\documentclass[elsarticle,twocolumn,english,superscriptaddress,aps,showpacs,floatfix,fleqn]{revtex4}

\usepackage{amsmath}
\usepackage{amsfonts}
\usepackage{amstext}
\usepackage{amssymb}
\usepackage{graphicx}
\usepackage{draftcopy}
\usepackage{flafter}
\usepackage{float}
\usepackage{tabularx}
\usepackage[latin1]{inputenc}
\usepackage{babel}

\graphicspath{{Figs/}}

\def\re#1{\mathcal{R}{e}#1}
\def\im#1{\mathcal{I}{m}#1}

\makeatother

\begin{document}

\title{Control of Majorana Edge Modes by a g-factor Engineered Nanowire
Spin Transistor}

\author{Amrit De}
\thanks{Corresponding author. Email : amritde@gmail.com}
\affiliation{Department of Physics \& Astronomy, University of California, Riverside,
California 92521, USA}

\author{Alexey A. Kovalev}
\affiliation{Department of Physics \& Astronomy, University of California, Riverside,
California 92521, USA}
\affiliation{Department of Physics \& Astronomy and Nebraska Center for Materials and Nanoscience, University of Nebraska-Lincoln,
Lincoln, NE 68588, USA}

\date{\today}
\begin{abstract}
We propose the manipulation of Majorana edge states via hybridization
and spin currents in a nanowire spin transistor. The spin transistor
is based on a heterostructure nanowire comprising of semiconductors
with large and small g-factors that form the topological
and non-topological regions respectively. The hybridization of bound edge states
results in spin currents and $4\pi$-periodic torques, as a function
of the relative magnetic field angle -- an effect which is dual to
the fractional Josephson effect. We establish relation between torques
and spin-currents in the non-topological region where the magnetic
field is almost zero and spin is conserved along the spin-orbit field
direction. The angular momentum transfer could be detected by sensitive
magnetic resonance force microscopy techniques.

\end{abstract}
\maketitle

\section{Introduction.}

It is believed that nanowires with strong spin-orbit interactions
can realize topologically protected quantum bits (qubits) \cite{Mourik:Science2012,Deng:NanoLett2012,Rokhinson:nov2012}
based on Majorana zero energy modes \cite{Majorana:1937,Wilczek:2009,Service:2011,Beenakker:2013,Alicea:2012}.
Some of the other proposals to realize topologically protected qubits
include schemes based on topological insulators \cite{Fu:2008,Nilsson:Sep2008},
fractional quantum Hall states \cite{Moore:1991}, cold atom systems
\cite{Gurarie:Jun2005,Jiang:Jun2011}, p -wave superconductors \cite{Rice:1995}
and superfluids in $^{3}\mbox{He-B}$ phase \cite{Silaev:2010}.

Typically proposals for observing these Majorana zero energy modes
are based on quantum tunneling and transport type phenomena \cite{Bagrets:Nov2012,Liu:Dec2012,Pikulin:2012,Lee:Oct2012,DasSarma:2012,Finck:Mar2013}.
Some exciting recent proposals for observing these edge modes are
based on the unconventional Josephson effect with a $4\pi$ periodicity
\cite{Kitaev:2001,Fu:Apr2009,Lutchyn:2010,Jiang:Nov2011}. A dual
effect has also been suggested in which case a torque between magnets
exhibits $4\pi$ periodicity in the field orientations \cite{Meng:PRB2012,Jiang:prb2013}.

It is the dual of Josephson effect that can in principle be employed
in spintronic devices. In particular, it is important to understand
the role of mechanical torques that should inevitably accompany Majorana
hybridization due to conservation of angular momentum. It has been
predicted that conservation of angular momentum in macrospin molecules
can result in quantum entanglement of a tunneling spin with mechanical
modes \cite{Kovalev:PRL2011,Garanin:PRX2011}. A flow of spin current
between two magnets has been demonstrated to induce spin-transfer
torque effect \cite{Berger1996,Slonczewski1996} and mechanical torques
\cite{Kovalev:Jan2007,Zolfagharkhani:dec2008}, also by conservation
of angular momentum. A flux qubit has been shown to decohere due to
exchange of angular momentum between the qubit and elastic modes of
a solid \cite{Chudnovsky:2012}.

\begin{figure}
\includegraphics[clip,width=0.4\textwidth]{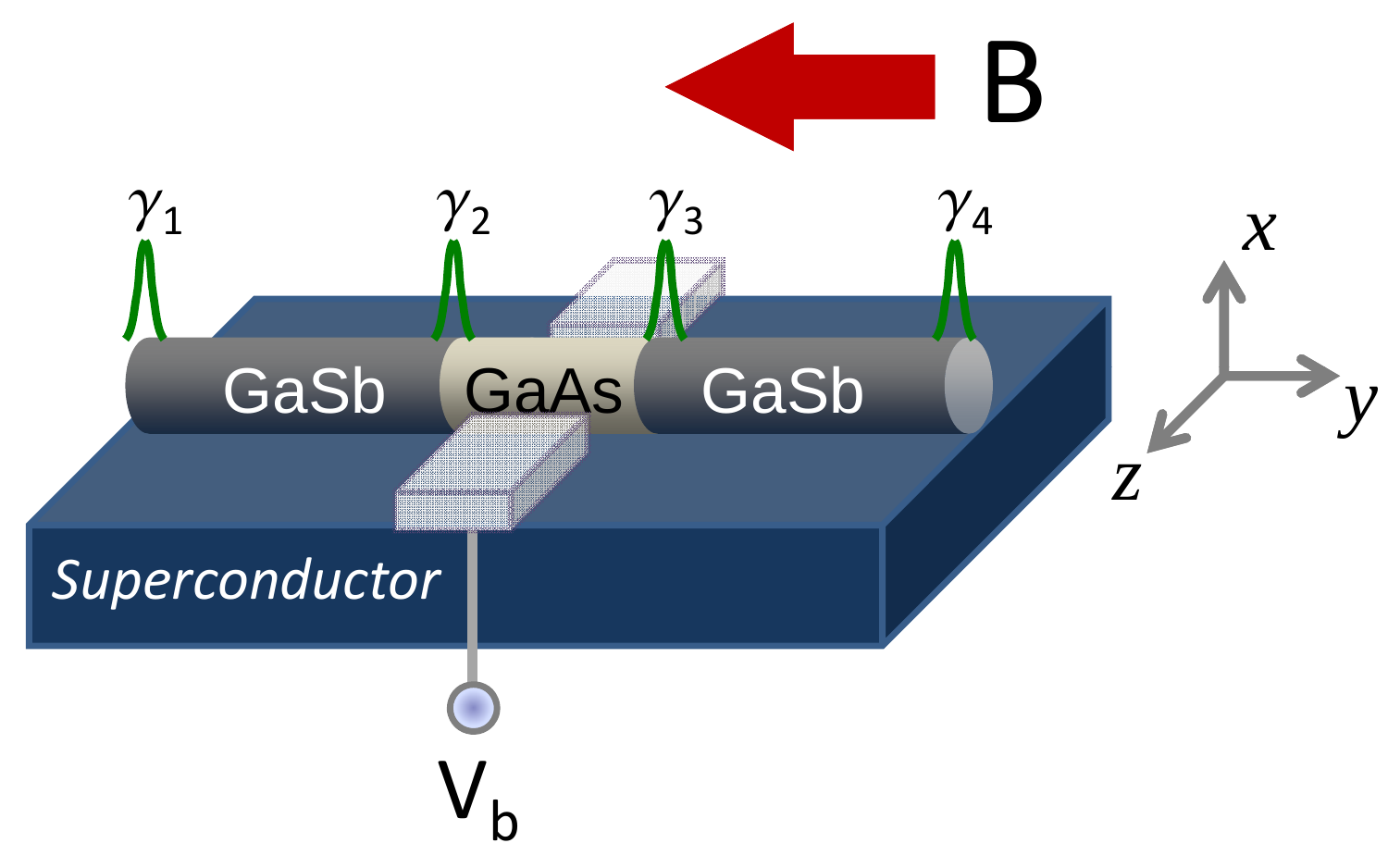} \caption{(Color online) A semiconductor nanowire with an embedded type-II GaSb-GaAs-GaSb
quantum well placed on top of s-wave superconductor in the presence
of a magnetic field. The relatively much smaller g-factor of GaAs
makes it a non-topological region. The gate voltage, $V_{b}$, can
be used in order to control the hybridization between the edge modes
formed at the hetero-junction interface. }

\label{figA}
\end{figure}

In this paper we propose the manipulation of Majorana edge states
via hybridization and spin currents using a gated nanowire spin transistor.
For our calculations we consider a spin transistor that comprises
of GaSb-GaAs-GaSb type-II quantum well with strong spin-orbit interactions.
Our choice of semiconductors is based on a number of factors. First,
the formation of topological and and non-topological region of the
wire is determined by the relative $g$-factors of the wires. While
GaSb has a large $g\approx-9$, the relatively smaller $g\approx-0.3$
for (GaAs) giving an excellent $g-$factor contrast of about $30$.
GaSb-GaAs nanowires and other Sb containing hetero-junction nanowires
have been grown \cite{Guo2006,Ganjipour2011,Dheeraj2008,Borg2013}.
Though InSb has a much higher $g$-factor and InSb-InAs heterojunction
nanowires have been grown\cite{Ercolani2009,Pitanti2011}, their $g-$factor
contrast is only about $4$. Second, the relatively small conduction
band offset of these two materials leads to the formation of a shallow
type-II quantum well, thereby only requiring a small gate voltage
to raise the chemical potential in the well region.

Typically strong quantum confinement effects can strongly alter the
electronic $g-$factors and drive it towards the bare electron value
due to effects of orbital angular momentum quenching \cite{Pryor2006b,De2009PRL}
in materials with sufficiently strong spin-orbit coupling. In the
case of InAs-InP nanowire quantum dots it has been shown that the
$g-$factor can be tuned through zero\cite{Bjork2005,De2007}. Therefore
in the case of GaAs (which has a small negative bulk $g$), a sufficiently
small quantum well should drive $g$ towards $0$, thus making it
a perfect non-topological region.

The topological region semiconductor should have strong spin-orbit
interactions and large Zeeman splitting. The schematic of our proposal
is shown in Fig.\ref{figA}. The wire is proximity-coupled to an $s$-wave
superconductor which results in proximity induced pairing in the wire.
The spin transistor allows the bound edge states to hybridize thus
resulting in spin-current induced $4\pi$-periodic torque, as a function
of the relative magnetic field angle. As an example of application
for our proposal, arrays of nanowires with zero energy edge modes
could be used as quantum memory -- in which case there arises a need
to efficiently read out information from memory elements. The spin
and angular momentum flows discussed here could be employed for that.
In general the nanowire architecture allows the combination of various
lattice mismatched materials and have attracted much attention due
to their potential electronic and optoelectronic applications such
as single electron transistors, field sensors, and low-power electronics
\cite{Thelander2003,Samuelson2004,Borg2013}.

We also establish relation between torques and spin-currents in the
non-topological region where the magnetic field is almost zero and
spin is conserved along the spin-orbit field direction. Sensitive
magnetic resonance force microscopy measurements can provide further
evidence for the existence of these edge modes and their hybridization.
Finally, we show that this non-dissipative spin current can be controlled
by the external gate voltage (see Fig. \ref{figA}) which leads to
similar functionality with Datta and Das spin-field-effect transistor
\cite{Datta:1990}.

\section{Tight Binding Calculations for Spin Currents and Edge hybridization}

Consider a semiconducting quantum wire, with Rashba spin-splitting, placed on top of a superconducting substrate (as per the coordinates shown in Fig. \ref{figA}). In this solid state system, Majorana fermions are charge less, localized zero-energy collective quasiparticle excitations of the superconducting ground state that satisfy the Bololiubov-de Gennes (BdG) Hamiltonian:
\begin{equation}
\begin{aligned}\mathcal{H}= & \frac{k^{2}}{2m}\tau_{z}+i\alpha_{so}k_{y}\tau_{z}\sigma_{z}+\boldsymbol{\Delta}'\cdot\boldsymbol{\tau}+{\bf B}\cdot\boldsymbol{\sigma}.\end{aligned}
\label{eq:BdG}
\end{equation}
where we have used the Nambu spinor basis $\Psi^{T}=(\psi_{\uparrow},\psi_{\downarrow},\psi_{\downarrow}^{\dagger},-\psi_{\uparrow}^{\dagger})$,
$\boldsymbol{\sigma}$ and $\boldsymbol{\tau}$ are Pauli vectors
that respectively act on particle and hole sectors. Here $\alpha_{so}$
is the strength of Rashba spin-splitting term, ${\bf B}=[B_{o}\cos{\theta},-B_{o}\sin{\theta},-B_{z}]$
is the magnetic field vector, $\boldsymbol{\Delta}'=[\Delta\cos{\phi},\Delta\sin{\phi},-\mu]$,
$\mu$ is the chemical potential and $\Delta e^{i\phi}$ is the superconducting
pairing potential. In general, the energy spectrum of the BdG Hamiltonian
supports gapped and gapless phases. The overall phase diagram is more
complicated than TI edge systems \cite{Meng:PRB2012,Jiang:prb2013}
due to the presence of the $k^{2}\tau_{z}$ term. Here, we limit our
consideration to the $\Delta^{2}>B_{z}^{2}$ part of the phase diagram
where the energy bands are always gapped. There are two gaped phases,
-- topological(T) for $\Delta^{2}-B_{z}^{2}<B_{o}^{2}-\mu^{2}$ and
non-topological(N) for $\Delta^{2}-B_{z}^{2}>B_{o}^{2}-\mu^{2}$,
separated by a quantum phase transition at $\Delta^{2}-B_{z}^{2}=B_{o}^{2}-\mu^{2}$.

The coupling between a magnetic field and the spin of an electron
is determined by the $g$-factor, which would therefore determine
whether the semiconductor is in the $N$ or $T$ phase. Hence, it is
possible to engineer a nanowire quantum well structure that can form
$T|N|T$ or $N|T|N$ regions even when placed in a uniform magnetic
field. For a $T|N|T$ type system, the hybridization across the $N$
region (which forms the well) can be gate controlled. Our proposed
spin transistor that comprises of GaSb-GaAs-GaSb type-II quantum well
within a nanowire is shown in Fig. \ref{figA}.

In order to treat arbitrary 1D heterostructures and non-uniform magnetic
fields, we transform the BdG Hamiltonian,~Eq. (\ref{eq:BdG}), onto the following tight
binding model with nearest neighbor hopping:
\begin{equation}
\begin{aligned}
\mathcal{H}= & \sum_{j,\sigma,\sigma'}\left[c_{j+1\sigma}^{\dagger}(-t_{0}\sigma_{0}+i\dfrac{\alpha_{j}}{2}\sigma_{z})_{\sigma\sigma'}c_{j\sigma'}+H.c.\right]\\
 & +\sum_{j,\sigma}(2t_{0}-\mu_{i})c_{j\sigma}^{\dagger}c_{j\sigma}+\sum_{j}(\widetilde{\Delta}_{i}c_{j\uparrow}^{\dagger}c_{j\downarrow}^{\dagger}+H.c.)\\
 & +\sum_{j}(\tilde{B}_{j}c_{j\uparrow}^{\dagger}c_{j\downarrow}+H.c.)\,
\end{aligned}
\label{eq:TBham}
\end{equation}
where we have used the complex parameters $\widetilde{\Delta}=\Delta\exp(i\phi)$, $\widetilde{B}=B_o\exp(-i\theta)$ and $c^\dagger_{j\sigma}(c_{j\sigma})$ creates (annihilates) an electron of spin $\sigma$ on site $j$. The proximity induced gap $\Delta=0.5meV$.
Here $t_{0}=\hbar/2m^{*}a^{2}$ is the hopping strength, $\alpha_{j}=\alpha_{so}^{(j)}/a$
where $a$ is the lattice spacing. In our calculations, we use the
following parameters for GaSb and GaAs, $m_{GaSb}=0.041~m_{e}$, $m_{GaAs}=0.067~m_{e}$,
$\alpha_{so}^{GaSb}=0.187~eV\textrm{\AA}$ and $\alpha_{so}^{GaAs}=0.024~eV\textrm{\AA}$.
The magnetic field at each site is given by $\tilde{B}_{j}=g_{j}\mu_{B}{\mathcal{B}}/2$,
where $\mu_{B}$ is the Bohr Magneton, $g_{j}$ is the Lande g-factor
of the semiconductor at that given lattice site and $\mathcal{B}$ is the applied field. Although quantum confinement
effects can alter the electronic $g-$factors in materials with sufficiently
strong spin-orbit coupling \cite{Pryor2006b,De2009PRL,De2007}, as
we are considering non-topological well regions that are fairly large
-- we use the bulk $g-$factors: $g_{GaAs}=-0.32$ and $g_{GaSb}=-8.72$.

\begin{figure}
\includegraphics[clip,width=0.5\textwidth]{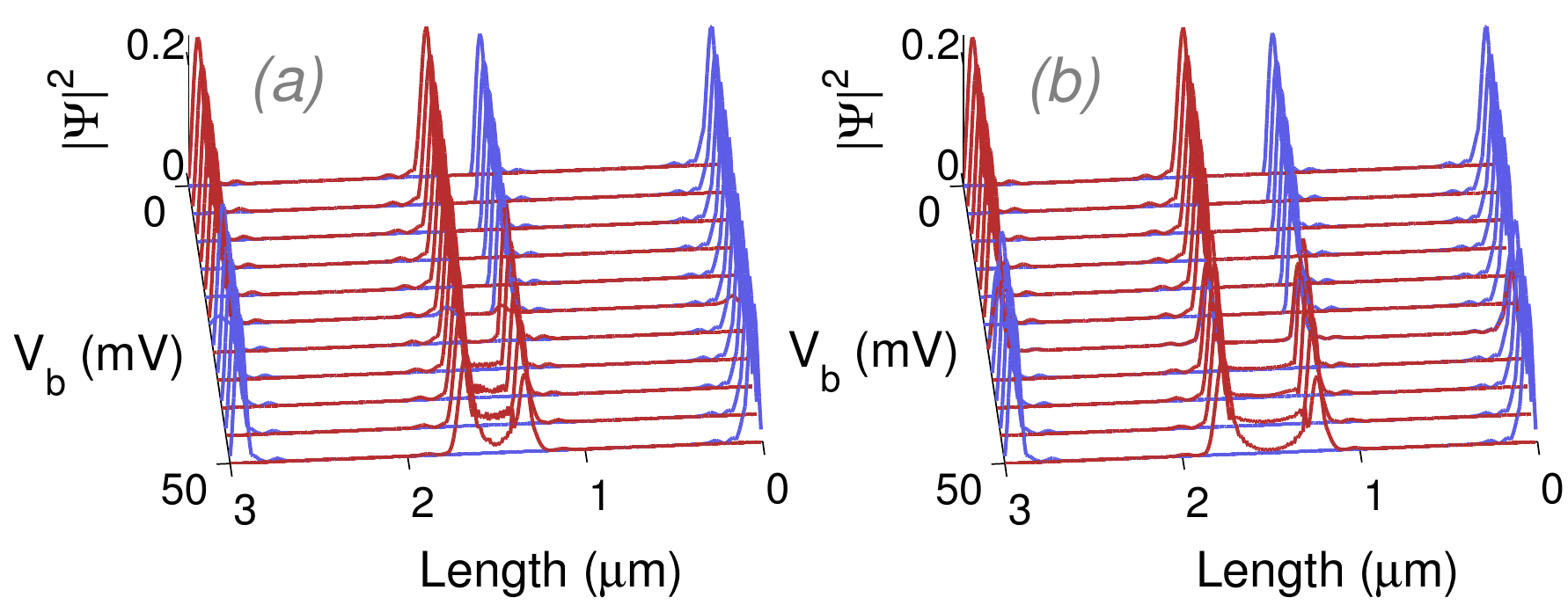} \caption{(Color online) Edge modes as a function of hetero-junction bias voltage
$V_{b}$ with the wire subjected to constant magnetic field for the
non-topological GaAs well widths of (a) 200 nm and (b) 400 nm. Hybridization
of the Majorana edge modes is seen at $V_{b}=30\, meV$.}
\label{Psi}
\end{figure}

The overall length of the wire was taken to be $3\mu\mbox{m}$ corresponding
to $300$ grid sites for a grid spacing of $a=10\mbox{nm}$. Typically
the GaSb section of the nanowires are about 60 nm in diameter, while
the GaAs sections are about 40 nm wide\cite{Ganjipour2011}. Taking
the effective masses, the transverse quantization, the bandgaps and
the valance band offset of these semiconductors in account, we estimate
that the barrier height of the quantum well is about $52$ meV.

\begin{figure}
\includegraphics[clip,width=0.48\textwidth]{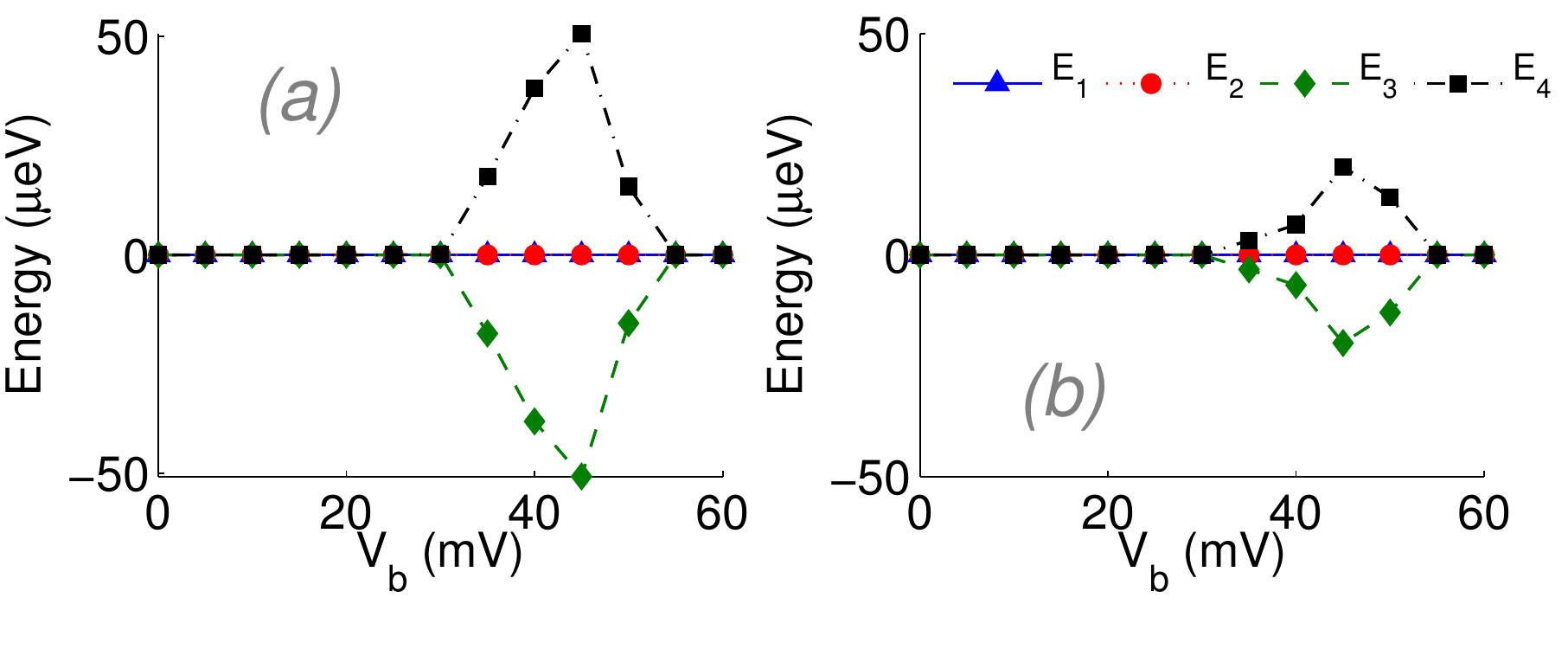} \caption{(Color online) The hybridization energies as a function of bias voltage
corresponding to Fig.\ref{Psi} for well widths of (a) 20 nm and (b)
40 nm.}
\label{Epsi}
\end{figure}

In Fig.\ref{Psi} we show the edge states from our tight binding calculations
as a function of the bias voltage applied to the non-topological GaAs
well region for two different well widths. The respective energies
of these states are shown in Fig.\ref{Epsi}. It is clearly seen that
the shorter non-topological well region results in higher hybridization
energies due to more dominant finite size effects. In the absence
of any bias, the quantum well prevents the hybridization of the edge
modes as indicated by the separate red and blue edge states. This
separation of the edge modes persists till the threshold $V_{b}\approx30$
mV is reached at which point a split in the energy spectrum is seen
due to hybridization of the edge modes. As $V_{b}$ is further increased,
the edge modes abruptly return to their unhybridized state as the
bias voltage now acts as a barrier preventing any tunneling effects.
The unhybridized Majorana edge modes ($\gamma_{1}$ and $\gamma_{4}$)
are formed at the ends of the structure and they have nearly zero
energy. The hybridized edge modes ($\gamma_{2}$ and $\gamma_{3}$)
are formed in the middle and they have non-zero energy due to finite
size effects. This non-zero energy edge mode can now be manipulated
by spin currents and magnetic field gradients.

In general the three components of the spin current operator are as
follows
\begin{eqnarray}
J_{j}(r)={\mathcal{R}}e\left(\Psi^{\dagger}(r)\left[\sigma_{j}\frac{\partial\hat{H}}{\partial{p}}+\frac{\partial\hat{H}}{\partial{p}}\sigma_{j}\right]\Psi(r)\right)\label{Jxyz}
\end{eqnarray}
where $j=x,y,z$ and $\Psi(r)$ is the position dependent Nambu spinor
which can be calculated by diagonalizing the tight binding Hamiltonian.
The spin currents were numerically calculated using finite differences
and the tight binding wavefunctions.

\begin{figure}
\includegraphics[clip,width=0.5\textwidth]{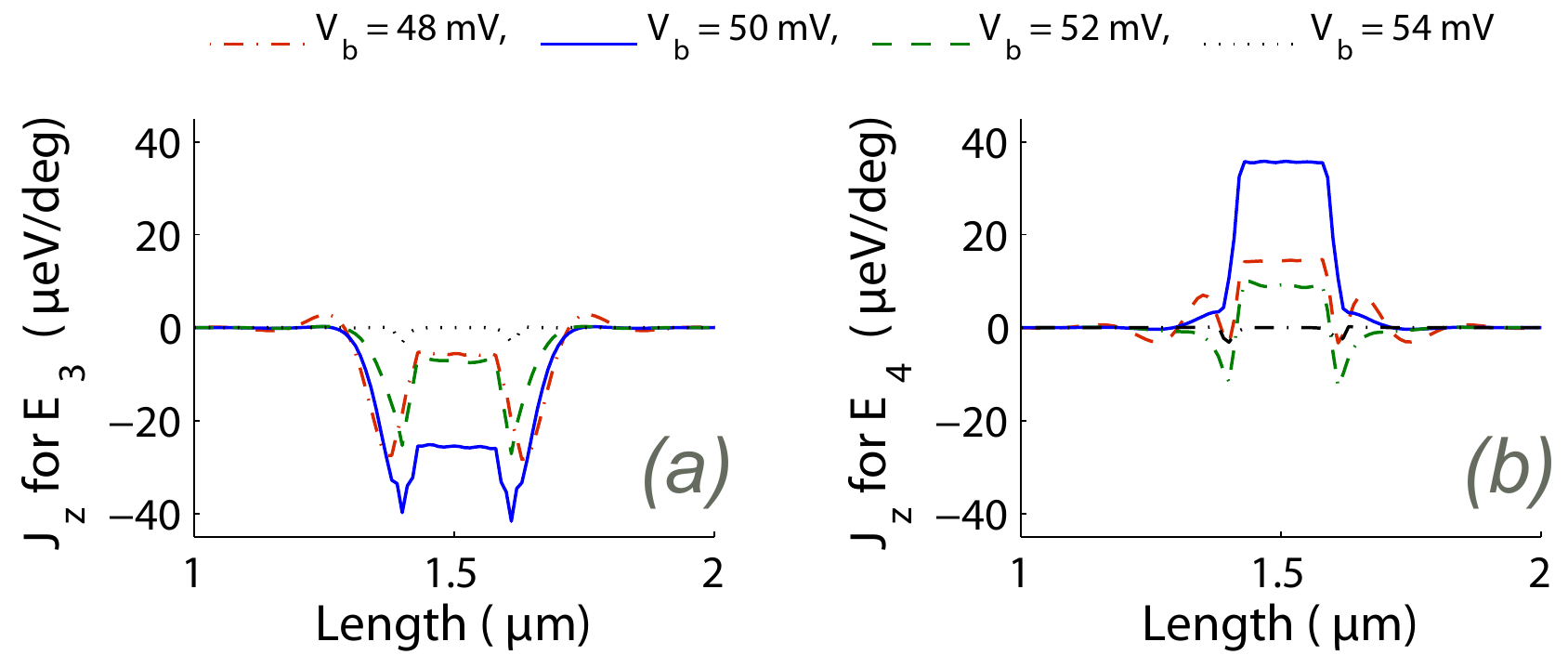} \caption{(Color online) The spatial distribution of the $z$ component of the
spin current for the (a) $E_{3}$ and (b) $E_{4}$ states as a function
of various bias voltages, $V_{b}$. The spin current only exists in
the middle section when the Majorana edge modes are allowed to hybridize
over the non-topological region. Note that the real part of $x$ and
$y$ components of the spin currents is zero. }
\label{Jy}
\end{figure}

The spatial distribution of the $z$ component of the spin current
for the hybridized edge states as a function of various gate bias
voltages is shown in Fig.\ref{Jy}. The Rashba spin-splitting is along
the $z-$direction and only the $J_{z}$ spin current is non zero
in the non-topological GaAs well region (spin is conserved along the
$z-$direction for non-topological region). It is seen that the gate
bias voltage can strongly affect the spin current. Overall the spin
current is nearly constant in the $N$ region and is therefore conserved.

\begin{figure}
\includegraphics[clip,width=0.48\textwidth]{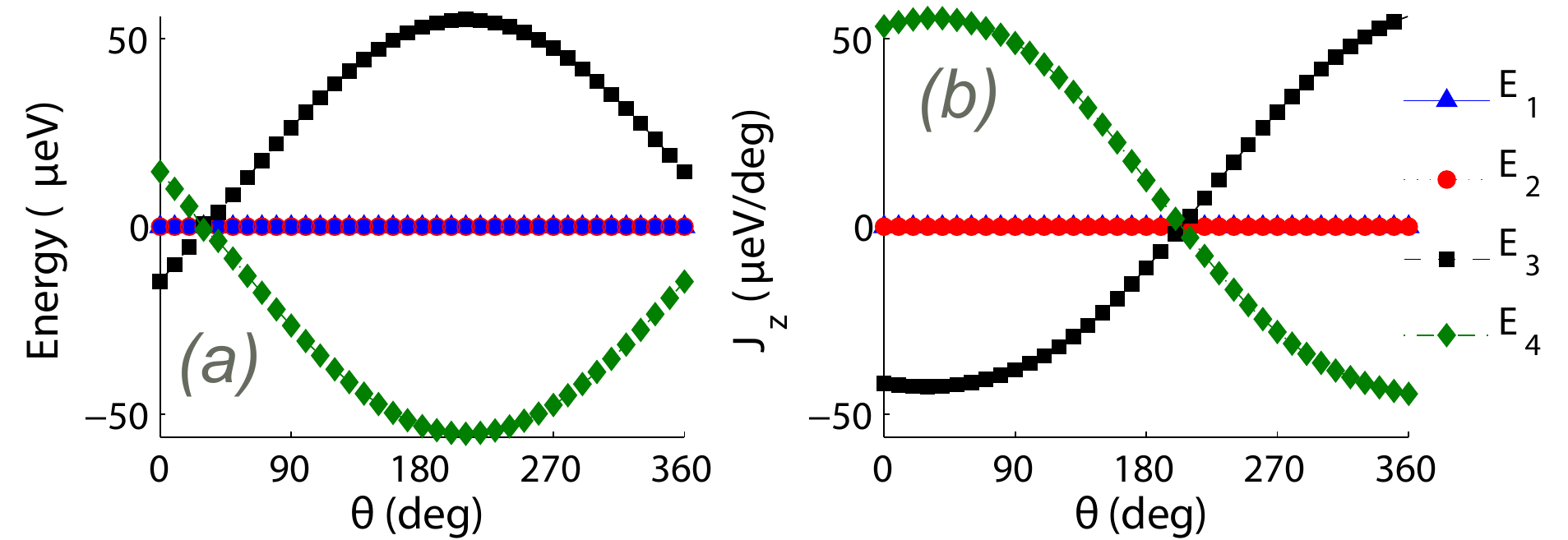} \caption{(Color online) (a) Hybridization energy of Majorana bound states as
a function of the relative angle, $\theta$, between the magnetic
fields on the left and right $T$ region (separate dipolar magnetic
tips with more localized fields can be used for this). Note the $4\pi$
periodic behavior. (b) The spin current as function of $\theta$ with
bias voltage of $V_{b}=50$ mV applied.}

\label{EAng}
\end{figure}

In Fig.\ref{EAng} (a), we plot hybridization energies of the edge
states as a function of the relative angle, $\theta$, between the
magnetic fields on the left and right topological GaSb region. Separate
magnetic tips with localized dipole type fields could be used to realize
this. Notice the $4\pi$ periodic behavior. The corresponding spin
torque $\partial E^{\mbox{n}}(\theta)/\partial\theta$ agrees with
spin current in Fig.\ref{EAng} (b). It is the $\theta$ dependence
of the hybridization energy that leads to mechanical torques. Such
torques could be detectable by the exertion of a mechanical torque
on a nano-magnetic tip \cite{Meng:PRB2012,Jiang:prb2013}. In the presence of hybridization in the topological region
we observe difference between the spin current and torque which
is the result of additional torques produced by the coupling between
the orbital and spin degrees of freedom.

\section{Analytic Results}

The numerical results presented in the previous section can be better
understood by the analytical results presented in this section.

In general, an effective two level system or a Majorana qubit can
be formed by four Majorana edges where three of the edges are hybridized
(in Fig. \ref{figA} we assume that edges $\gamma_{1}$ and $\gamma_{2}$
are separated from each other). The effective low energy Hamiltonian
can be written as
\begin{equation}
\mathcal{H}=iE^{\mbox{n}}(\theta)\gamma_{2}\gamma_{3}+iE^{\mbox{t}}\gamma_{3}\gamma_{4}\label{eq:rotations}
\end{equation}
where $\theta$ is the angle between the magnetic fields, $\gamma_{i}$
describe Majorana edge states that can form two fermions $b'=\gamma_{1}+i\gamma_{2}$
and $b''=\gamma_{3}+i\gamma_{4}$, where $b^{\dagger}(b)$ are ferminoic
creation(anhilation) operators.

We analyze the hybridization of the edge modes and resulting spin-currents
and torques on $T|N|T$ structures. In order to calculate $E^{\mbox{n}}(\theta)$,
we consider a semiconductor nanowire with two infinite $T$ regions
(GaSb) separated by a finite $N$ region (GaAs). We introduce parameters
$\{\Delta_{L},B_{L},\mu_{L},\theta_{L}\}$ for the left $T$ region,
$\{\Delta_{M},\mu_{M}\}$ for the middle $N$ region and $\{\Delta_{R},B_{R},\mu_{R},\theta_{R}\}$
for the right $T$ region. The phase of superconducting pairing is
assumed constant (i.e. $\phi=0$) throughout the wire. In order to
determine the bound state at a single $T|N$ boundary, we need to
find the $4-$component zero energy solution to the Hamiltonian (Eq.
\ref{eq:BdG}). We use ansatz $\Psi(x)=\exp(\kappa x)\Psi(\kappa)$,
where $\kappa$ is complex. In doing so we arrive at four solutions
that decay into the topological region, i.e. with $\re(\kappa)>0$,
and four solutions that decay into the non-topological region, i.e.
with $\re(\kappa)<0$. A linear combination of these solutions
on each side has to be continuous and have continuous derivative at
the $T|N$ boundary which leads to unique solution for the edge state.

We denote such solutions as $\left|\psi_{L}\right\rangle =\exp{[i\theta_{L}\sigma_{z}/2]}\left|\psi_{L}^{0}\right\rangle $
for the left Majorana edge and as $\left|\psi_{R}\right\rangle =\exp{[i\theta_{R}\sigma_{z}/2]}\left|\psi_{R}^{0}\right\rangle $
for the right Majorana edge. Next we employ lowest order perturbation
theory to find the hybridization energy of Majorana modes and spin
current at the boundary when the solutions for the left and right
edges weakly overlap. For the hybridization energy we obtain:
%\begin{equation}
%E^{\mbox{n}}(\theta)\approx\dfrac{\left|\left\langle \psi_{L}^{0}e^{-i\frac{\theta_{L}}{2}\hat{\sigma}^{z}}\right|H\left|e^{i\frac{\theta_{R}}{2}\hat{\sigma}^{z}}\psi_{R}^{0}\right\rangle \right|}{\sqrt{\left\langle \psi_{L}^{0}|\psi_{L}^{0}\right\rangle \left\langle \psi_{R}^{0}|\psi_{R}^{0}\right\rangle }},
%\end{equation}
\begin{equation}
E^{\mbox{n}}(\theta)\approx\dfrac{\left|\left\langle \psi_{L}\right|H\left|\psi_{R}\right\rangle \right|}{\sqrt{\left\langle \psi_{L}^{0}|\psi_{L}^{0}\right\rangle \left\langle \psi_{R}^{0}|\psi_{R}^{0}\right\rangle }},
\end{equation}
which becomes
\begin{equation}
E^{\mbox{n}}(\theta)\approx{E_{0}^{\mbox{n}}}\exp[{-\re(\kappa_{2}^{n})\ell_n}]
\cos\left[\dfrac{\theta}{2}+\Phi_{0}+\im(\kappa_{2}^{n})\ell_n\right],\label{eq:HybNT}
\end{equation}
where $\kappa_{2}^{n}=m^{*}/\hbar\bigl(i\alpha_{so}-i\sqrt{2(i\Delta+\mu)\hbar/m^{*}+\alpha_{so}^{2}}\bigr)$,
$E_{0}^{\mbox{n}}$ and $\Phi_{0}$ depend on the parameters of the $T$
and $N$ regions and not on the wire length, $\ell_{n}$, and $\theta$.

Similarly for a semiconductor nanowire with two infinite $N$ regions
separated by a finite $T$ region (GaAs), the hybridization energy
for the topological region is:
\begin{eqnarray}
{E^{\mbox{t}}} \approx {E_{0}^{\mbox{t}}}&\Bigl(&\exp[{-\kappa_{2}^{t}\ell_t}] + \\\nonumber
&&\left|A_{0}\right|\exp[{-\re(\kappa_{1}^{t})\ell_t}]\cos\left[\arg A_{0}+\im(\kappa_{1}^{t})\ell_t\right]\Bigr),\label{eq:HybT}
\end{eqnarray}
where $\kappa_{1}^{t}$ and $\kappa_{2}^{t}$ are the roots of $\sqrt{B^{2}-[\kappa^{2}(\hbar/2m)^{2}+\mu]^{2}}=\Delta+\alpha_{so}\kappa$
such that $\Re(\kappa)>0$. Here $E_{0}^{\mbox{t}}$ and $A_{0}$
depends on parameters of the $T$ and $N$ regions and not on $\ell_{t}$ and $\theta$.

From the perturbative solutions we can express spin current at point $r$ as
\begin{eqnarray}
J_{z}(r)=~~~~~~~~~~~~~~~~~~~~~~~~~~~~~~~~~~~~~~~~~~~~~~~~~~~~~~~\\
\re\left(\dfrac{\left[\psi_{L}^{\dagger}(r)\pm i\psi_{R}^{\dagger}(r)\right]\{\sigma_{z},\hat{\upsilon}\}\Bigl[\psi_{L}(r)\pm i\psi_{R}(r)\Bigr]}{2\sqrt{\left\langle \psi_{L}^{0}|\psi_{L}^{0}\right\rangle \left\langle \psi_{R}^{0}|\psi_{R}^{0}\right\rangle }}\right)\nonumber
\end{eqnarray}
where  $\hat{\upsilon}=\partial\hat{H}/\partial p$ and $\theta=\theta_{R}-\theta_{L}$.
This leads to:
\begin{equation}
J_{z}=\pm\dfrac{\partial E^{\mbox{n}}(\theta)}{\partial\theta},
\end{equation}
which relates spin current to the hybridization energy over the $N$
region. This shows that the torque $\partial E^{\mbox{n}}(\theta)/\partial\theta$
is generated solely by the spin current passing through the middle
$N$ region.

\section{Manipulation of Majorana qubit by gates}

Without loss of generality we can fix the electron parity, e.g. $1$.
The available Hilbert space of two fermions corresponding to four
edges ($b'=\gamma_{1}+i\gamma_{2}$ and $b''=\gamma_{3}+i\gamma_{4}$)
is $\alpha\left|1,0\right\rangle +\beta\left|0,1\right\rangle $.
This is equivalent to a Hilbert space of a spin $1/2$ system. The
Hamiltonian in Eq. (\ref{eq:rotations}) then becomes:
\begin{equation}
\mathcal{H}=\dfrac{E^{\mbox{n}}(\theta)}{4}\sigma_{x}+\dfrac{E^{\mbox{t}}}{4}\sigma_{z},\label{eq:Rabi1}
\end{equation}
which shows that by gate tuning $E^{\mbox{n}}$ and $E^{\mbox{t}}$
one can perform arbitrary rotations of the Majorana qubit. The relative
angle between magnetic fields, $\theta$, can be also used for manipulations.
Form the analysis in the previous section it becomes clear that rotations
along the $x-$axis are accompanied by mechanical torques. Nevertheless,
we estimate that this should not be a strong source of decoherence
\cite{Chudnovsky:2012}.

\section{Summary}

We have proposed the manipulation of Majorana edge states in a gated
nanowire spin transistor that comprises of GaSb-GaAs-GaSb type-II
quantum well. The formation of topological and and non-topological
region of the wire is determined by the large $g$-factor contrast
of the two semiconductors. In general larger spin-orbit interactions
(coupled with smaller band gaps) lead to larger g-factors. It is possible
to obtain a full Bloch sphere rotations of the Majorana qubit by gates
controlling the hybridization energies. The setup can be easily generalized
to a larger number interchanging topological/nontopological regions
where universal rotations can be achieved by applying time dependent
gates. Spin currents and torques can be further used in order to couple
such a wire to various read out schemes\cite{Kovalev:unpublished2013}. The flows of angular momentum,
comprising a signature of Majorana edge states, can be detected by
sensitive magnetic resonance force microscopy techniques.

\bibliographystyle{apsrev}
%\bibliography{bibliography,Mwire2}

%==============================================================================================================================

\end{document}